\documentclass[twocolumn,showpacs,preprintnumbers,amsmath,amssymb, prl]{revtex4}
\usepackage{graphicx}

\begin{document}
\title{THERMODYNAMIC PROPERTIES OF THE DISCOMMENSURATION POINT\\
 FOR INCOMMENSURATE
STRUCTURES:\\ A "THIRD-ORDER" PHASE TRANSITION}
\author{\normalsize A.N. Rubtsov$^a$ and T. Janssen$^b$}

\affiliation {$^a$ Physics Department, Moscow State University, Moscow, Russia\\
$^b$ Institute for Theoretical Physics, University of Nijmegen,
Nijmegen, The Netherlands}

\email{AR: alex@shg.ru, TJ: ted@sci.kun.nl}


\begin{abstract}
The consequences of the opening of a phason gap in incommensurate
systems are studied on a simple model, the discrete frustrated
$\phi^4$-model. Analytical considerations and numerical results
show that there is a very weak phase transition that can be
characterized as third order.
\end{abstract}

\maketitle

\section{Introduction}
It is well known that the modulation function of an incommensurate modulated
phase is smooth at the transition from disordered to incommensurate, and
that it may become discontinuous further from the transition line. In the incommensurate
phase it marks the transition from a simple modulation to a structure that is
locally like the commensurate structure that usually appears at another
phase transition from incommensurate to commensurate modulation. The phenomenon
of the appearance of a discontinuity is often called the transition by breaking
of analyticity [1] or the discommensuration transition [2], the latter because
at the transition the structure is locally periodic but with domain walls, called
discommensuration. The transition is accompanied by an opening of a gap
in the phason frequencies [3]. It has been studied on several models [2-9],
like the discrete frustrated $\phi^4$-model (DIFFOUR), the Frenkel-Kontorova model
and a double chain model for incommensurate composites.

From these studies the qualitative change in the modulation and the phason
gap opening have been well established. It is not so clear whether the transition
is a transition in the thermodynamic sense, and whether there are discontinuities
in the thermodynamic quantities. To that end we analyze the situation for
a displacively modulated structure using the DIFFOUR model. First the
ground state properties are analyzed. Then temperature effects are studied. The
analytical results are checked with numerical calculations.

\section{ Definitions and ground-state properties of the model}

    The model under consideration is the 3D DIFFOUR array.
It is a cubic lattice of 2-4 anharmonic double well oscillators, with
a harmonic coupling.
Each atom is coupled with 6 nearest neighbours and 2 next-nearest
neighbours in the $z$-direction. The potential energy of the model
is expressed as
\begin{widetext}
\begin{equation}\label{V0}
  V=-\frac{A}{2} \sum_n x_n^2 + \frac{B}{4} \sum_n x_n^4 +
  C\sum'_{n, m} (x_n-x_m)^2 +
  D\sum''_{n, m} (x_n-x_m)^2.
\end{equation}

%\end{widetext}

Here $A, B, C, D$ are constants. The third sum (marked with a
prime) runs over all pairs of the nearest neighbours $n$ and $m$,
and the last sum (with a double-prime) is over the pairs of the
next-nearest neighbours in the $z$-direction.

The ground-state properties of the model depend on the two
dimensionless combinations: $a=-A/C$ and $d=D/C$. Along a line in
the $a-d$ plane, the modulated phase with an incommensurate
modulation period becomes energetically favourable:
\begin{equation}\label{def}
    x_n=f(z),
\end{equation}
where $f$ is a periodic function with, generally, an irrational
period. For a small $a$, $f(z)$ is a sinus-like continuous
function. With an increase of $a$ the modulation function becomes
less smooth and at certain temperature can loss the continuity
("breaking of analyticity" or discommensuration transition). The
part of the ground-state "phase diagram", containing the
incommensurate phase, is schematically shown in Fig.1. Typical
profiles of the modulation function are shown in Fig. 2. The
plateaus at the discontinuous curve correspond to a locally
periodic structure with discommensuration.

One can note that even the discontinuous modulation function
roughly resembles the graph of sinus.
Therefore it is reasonable first to consider a sinusoidal trial modulation $f_0(z)=x_0 \sin k z$. The
minimisation of the potential
energy gives $\cos k=-\frac{C}{4 D}$, $x_0^2=\frac{4}{3} (A-\frac{C^2}{2 D}- 4 C
- 8 D)$.

The following procedure allows to describe qualitatively the
appearance of the discontinuities. Let us use $f_0(z)$ as a
starting point, and find a function $f_1(z)$ that minimises the
potential energy of each layer in the fixed field of its
neighbours:

%\begin{widetext}
\begin{eqnarray}\label {f1}
  V_1(z)\equiv \frac{B f_1^4(z)}{4} + (-A+4 C+ 4 D ) \frac{f_1^2(z)}{2} %     \nonumber
   - 2 f_1 (z)
  (C f_0(z+1) + C f_0(z-1) + D f_0(z+2) + D f_0(z-2))
  = \min.
\end{eqnarray}
\end{widetext}

\begin{figure}
\includegraphics[width=8cm]{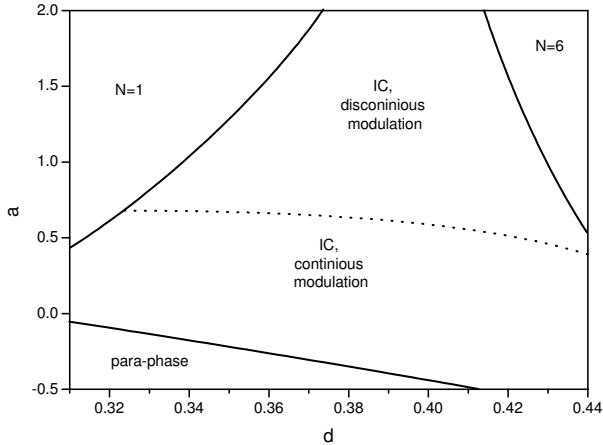}
\caption{\label{fig:epsart} A part of the ground-state phase
diagram for the DIFFOUR model (1), including the incommensurate
(IC) phase. Regions denoted by N=1 and N=6 correspond respectively
to the (commensurate) ferroelectric phase and to the modulated
phase with a integer period of 6 layers.}
\end{figure}

\begin{figure}
\includegraphics[width=8cm]{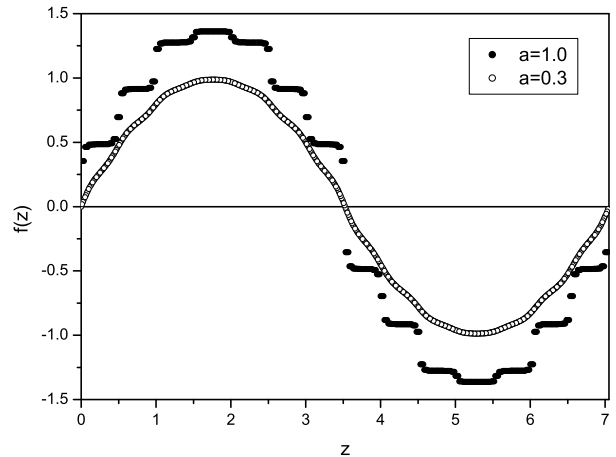}
\caption{\label{fig:epsart} Examples of the ground-state
modulation function above ($a=0.3$) and below ($a=1.0$) the
breaking of continuity transition. Both curves are plotted for
$d=-0.4$.}
\end{figure}

One can consider this expression as the first iteration in the
determination of the modulation function for a fixed  period of
modulation. However, we shall consider only the first iteration.
Physically, this is a condition for the ground-state state of an
anharmonic oscillator in the given external field formed by the
neighbouring particles. The value of the field passes through zero
at $z=0$ and at $z=\pi/k$. The qualitative behaviour of $f_1(z)$
is determined by the sign of $-A+4 C+ 4 D $. If this combination
is positive, $f_1(z)$ is a continuous function. Otherwise,
discontinuous steps from $-\sqrt{A-4 C-4 D}$ to $\sqrt{A-4 C-4 D}$
appear at $z=0$ and at $z=l/2$.

The situation at the points of discontinuity resembles quite much
the Landau scenario of second-order phase transitions. Thus
the contribution to the ground-state energy {\it from these two points} is
of the Landau type - for example,
it shows a discontinuity in the second derivative with respect to $A$.
However, to obtain the {\it total} energy, one should
integrate over the modulation period. It is
useful to consider the derivative of the ground-state energy:
\begin{equation}
  \frac{d V_{gs}}{d A}=\frac{\partial V_{gs}}{\partial A} \propto \int \frac{f^2(z)}{2} dz
\end{equation}
To establish this formula we took into account that
$\frac{\partial V_{gs}}{\partial x_m}=0$, as $V_{gs}$ should be
minimised with respect to $x_m$.
The behaviour in the transition point with $f=f_1(z)$ can be analysed by means of the
expansion in powers of $(-A+4 C+ 4 D)$. It turns out that both $d V_{gs}/{d
A}$ and $d^2 V_{gs}/{d A^2}$ are continuous,
and only the third derivative shows a step.

Quantitatively, the approach described above is very inaccurate.
For example, for $D=-0.4C$ it predicts the breaking of continuity at
$A = 2.4 C$, whereas actually it takes place at $A\approx 0.6
C$. One can increase the accuracy, considering instead of
(\ref{f1}) the minimum of each three neighbouring layers in
a given field of the surrounding layers:
\begin{eqnarray}
\nonumber
  &&V_3\equiv
  (f_1^4(z)+f_1^4(z+1)+f_1^4(z-1))/{4} +\\ \nonumber
  && (-A+4 C+ 4 D ) (f_1^2(z)+f_1^2(z+1)+f_1^2(z-1)) /2  \\       \nonumber
  && - 2C f_1(z) f_1 (z+1) - 2C f_1(z) f_1 (z-1) \\ \nonumber
  &&-2D f_1(z-1) f_1 (z+1) \\
  &&- 2 D f_1 (z) (f_0(z+2) + f_0(z-2))\\       \nonumber
  && - 2 f_1(z+1)(C f_0(z+2)+D
  f_0(z+3)) \\ \nonumber
  &&- 2 f_1(z-1)(C f_0(z-2)+D f_0(z-3))
  = \min.
\end{eqnarray}
The function $f_1$ becomes discontinuous when the minimal eigenvalue of
the $3 \times 3$ matrix of the second derivatives of $V_3(0)$
becomes zero. This gives the following condition for the breaking of
continuity: $-A + 4 C + 3 D - \sqrt{8 + (D - 3 f_1^2(1))^2)}
+3 f_1^2(1)/2=0$. The prediction for the $D=-0.4C$ system is now
much better: $ A\approx 1.0 C$.

It  should be noted also that the continuity of the modulation
function actually breaks down not at the single point $z=0$.
Indeed, the exactly calculated modulation function is a fixed
point of the transformation (\ref{f1}). It is clear from this
transformation, that if there is a discontinuity at certain $z$,
the discontinuity at $z \pm 1$ should be present also. Therefore,
in the mathematical sense, $f(z)$ is discontinuous in all points
when the continuity is broken in a single point. Practically (see
Fig.2), the "parent" discontinuity at $z=0, \pi/k$ is the most
remarkable, the magnitude of "satellite" discontinuities falls
down fast.

Although the above approach is very simplified, it clearly shows
the physical picture of the breaking of continuity. The
qualitative prediction about the continuity of $d^2 V_{gs}/{d
A^2}$ seems to be correct as well. Fig. 3 shows a numerical result
for this quantity, which is a smooth line.

At the same time, it is worth to note that the "breaking of
continuity" affects the spectrum of elementary excitations
essentially. For a continuous modulation function, a phason branch
is present in the spectrum [2, 3, 9]. It shows a linear
(phonon-like) dispersion, with a phase velocity
$v\propto\left(\int_0^l \left({\partial f}/{\partial z}\right)^2
dz \right)^{-1/2}$. The zero-frequency Goldstone mode corresponds
to the uniform shift of the modulation phase; its  eigenfunction
is $u_0\propto v {\partial f}/{\partial z}$ (the normalisation is
$\int u^2 dz=1$). For a discontinuous $f(z)$, the integral for $v$
diverges, indicating the presence of a gap at the beginning of the
spectrum. This regime can be described as follows. Formally, one
can still calculate $u_0$ from the above formula. For a modulation
function (\ref{f1}), one obtains that $u_0$ becomes the two
singular peaks at $z=0, \pi/k$. Physically, this means that the
low-lying excitations are the vibrations of atoms only in the
points where the modulation is broken. Therefore, their frequency
is determined by the second derivative of $V_1$, $\omega_0^2
\propto d^2 V_1/dz^2$. One can interpret these vibrations as
domain wall oscillations. Consequently, the presented model
predicts a gap, which is proportional to the square of the
discontinuity of the modulation.

\begin{figure}
\includegraphics[width=8cm]{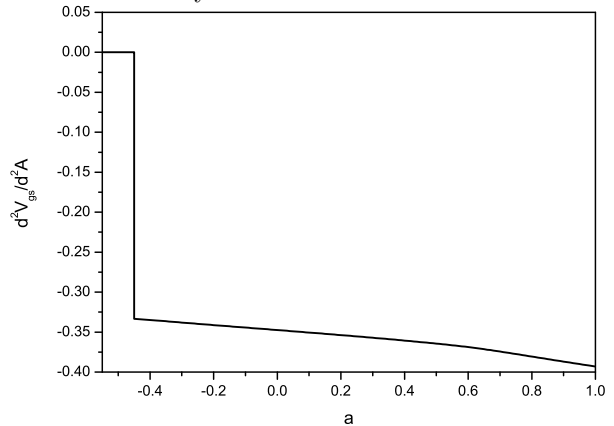}
\caption{\label{fig:epsart} The second derivative of the
ground-state potential energy with respect to $a$ for the system
with $d=-0.4$. There is a step at $a=-0.45$, corresponding to the
transition from para-phase to the incommensurate state. No
singularities at the breaking of continuity point ($a\approx0.6$)
are observed.}
\end{figure}

\section{ Finite temperature: the mean-field approach and Monte Carlo results}

Let us turn to the case of finite temperature $T$.
The partition function is equal to
\begin{equation}
  Z=\int \exp(-V/T) dx_1 ... dx_n
\end{equation}
(Boltzmann constant is equal to unity).

In the mean-field consideration, the inter-atomic correlations are
neglected, so that the average position of an on-site oscillator
$\bar{x}(z)$
is given by a self-consistent equation
\begin{widetext}
\begin{eqnarray}\label{MF}
  \bar{x}(z)&=& \bar{x}(f(z))= \frac{\partial F (f(z))}{\partial f}\\
  \nonumber
  F (f(z)) &\equiv& -T \ln \int \exp\left(-\frac{1}{T}\left(\frac{x^4}{4}+\frac{(-A+12 C+4 D) x^2}{2}-f(z)
x\right)\right)
  dx;\\   \nonumber
  f(z)&=&2 C (4 \bar{x}(z)+\bar{x}(z-1)+\bar{x}(z+1))+2 D
  (\bar{x}(z-2)+\bar{x}(z+2)).
\end{eqnarray}
Here $f(z)$ is an average force acting on the oscillator from its
neighbours.

Note that these equations determine the one-to-one correspondence
between $\bar{x}(z)$ and $f(z)$, so that one can construct a
function $f(\bar{x})$. It can be checked that equations (\ref{MF})
are the minimum conditions for the function
\begin{eqnarray}\label{FMF}
  F_{MF}&=&\sum_n \left(F(f(x_n)-F(0)+x_n f(x_n)-(6 C+2 D) x_n^2) \right)
  +\\     \nonumber
  &&C\sum_{n,m}' (x_n-x_m)^2 +  D\sum_{n,m}'' (x_n-x_m)^2.
\end{eqnarray}
%\end{widetext}
Thus in the mean-field approximation the problem at a finite temperature is
replaced by the problem for the ground-state on an effective
system (\ref{FMF}). The on-site potential of this effective system depends
on temperature. The value of $F_{MF}$ can be interpreted as a
mean-field expression for the free energy of the system. As the
temperature goes to zero, (\ref{FMF}) tends to (\ref{V0}), if $A$
is not too large ($A<12 C+ 4D$).

The breaking of continuity of the modulation function occurs at a
certain critical temperature $T_{c}$.
The considerations from the previous section can
directly be applied here. In particular, the heat capacity, being the
second derivative of the free energy, does not show
a discontinuity at $T_c$.

To go beyond the mean-field approach, it is necessary
to consider the role of the fluctuations.
Here we discuss, if the fluctuation phenomena can destroy the
phase with a broken continuity.
As it is demonstrated, the breaking of continuity occurs due to the
appearance of the effective double-well potential for the layer
with $z=0$ (formula (\ref{f1})). To incorporate the fluctuation in this model,
it is natural to suppose that in the zeroth approximation the modulation
function is sinusoidal and fluctuations are absent. Instead of (\ref{f1}) consider
the statistical sub-ensemble with energy
%\begin{widetext}
\begin{eqnarray}
   \sum_i \frac{x_i^4}{4} + \frac{(-A+4 C+ 4 D ) x_i^2}{2} +
   \sum'_{i,j} C (x_i-x_j)^2 %\\       \nonumber
   - 2 \sum_i x_i
  (C f_0(z+1) + C f_0(z-1) + D f_0(z+2) + D f_0(z-2))
\end{eqnarray}
\end{widetext}
Here the indices run over the layer with a given $z$-coordinate.
The layer is in fact the 2D discrete $\phi^4$ model [10]
placed in an external field. The dependence of the
average of $x_i$ on the value of the external field is
discontinuous, if the layer without the external field
shows a phase transition. Since a phase transition  occurs
in  the 2D discrete $\phi^4$ model indeed,
we conclude that fluctuations
do not destroy the discontinuity of the modulation function for the model
under consideration. It can be noted that for a 2D DIFFOUR model
this is not true. In fact, in this case we would come to the 1D discrete
$\phi^4$ chain, which does not show a phase transition.

To check the conclusion about the thermal behaviour of the heat
capacity, we performed a Monte Carlo simulation of the system with
a modulated ground state with a discontinuous modulation function.
The system was studied for $a=1.5, d=-0.4$. A slab of $8 \times  8
\times 557$ atoms with periodic boundary conditions and the
standard Metropolis algorithm was used. Initially we put the
system in its ground-state, and set the temperature to be small.
Then the temperature was slowly increased (about $3 \cdot 10^9$
Monte Carlo trial steps were used for each value of temperature).
The average potential energy $<E>$ was calculated at each point.
Afterwards, the thermal capacity $d<E>/dT$ was obtained. Thermal
behaviour of this quantity is presented in Fig. 4. There is a
clear peak at $T B/C^2\approx 3.7$ corresponding to the transition
between the modulated and disordered phase. The breaking of
continuity of the modulation function occurs at a lower
temperature, but it does not result in any remarkable feature of
$d\langle E(T)\rangle /dT$.

\begin{figure}
\includegraphics[width=8cm]{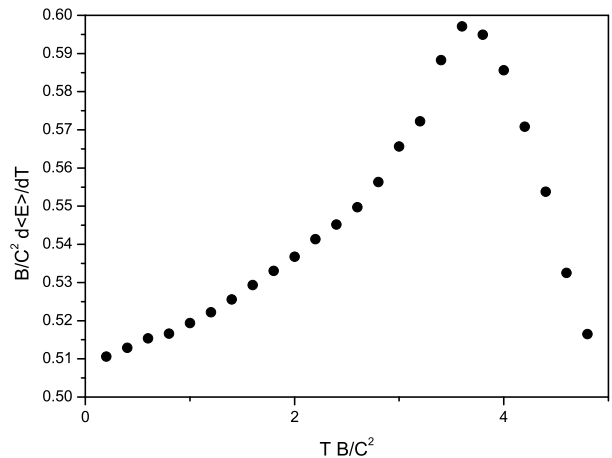}
\caption{\label{fig:epsart} Monte Carlo result for the heat
capacity of the system $a=1.5, d=-0.4$. The ground state of this
system is incommensurate with a broken continuity. The data shows
a clear peak at the transition to the disordered phase ($T B/
C^2\approx 3.7$) and no other pronounced features.}
\end{figure}

\section{Concluding remarks}

In several model systems for aperiodic crystals there is a transition from a smooth modulation function to a
modulation function with discontinuities. This transition has been
called the discommensuration transition of the transition by breaking of
analyticity or continuity. In the phase space region where the modulation
function is continuous there is in these models a phason excitation of zero
frequency. In the other region there is a phason gap. However, neither the
structural transition nor the opening of a phason gap has been observed in
experiment. There are even few systems where a gap-less phason or phason with very low
frequency has been found. Generally this is attributed to pinning by defects,
or other friction mechanisms [11]. Another possible method to check the existence
of the transition is by measuring thermodynamic quantities. In the model studied
here, the DIFFOUR model, the transition is very weak. It can be called to be of
third order. The conclusion is that the transition could only be observed
by a precise structure determination. However, in general the discontinuity will
be not strong enough to be seen.

\vspace{1.cm}

\noindent {\bf Acknowledgment}

The work was supported by an INCO Copernicus program (Grant ERBICI
15 CT 970712).

\vspace{1.cm}

\noindent {\bf References}

\noindent\begin{itemize}
\item[1] S. Aubry, in {\it Solitons and condensed matter physics}, (eds. A.R. Bishop and T. Schneider), Springer,
Sol. St. Sc. {\bf 8}, 264 (1978).
\item[2] T. Janssen and J.A. Tjon, Phys. Rev. B{\bf 25}, 3767 (1982).
\item[3] T. Janssen and J.A. Tjon, J.Phys.C {\bf 16},4789 (1983).
\item[4] F. Borgonovi and D. Shepelyanski, Europhys. Lett. {\bf 21}, 413 (1993).
\item[5] T. Kawaguchi and H. Matsukawa, Phys.Rev. B {\bf 56}, 13932 (1997), Phys. Rev. B {\bf 58}, 15866 (1998).
\item[6] A.M. Morgante, M. Johansson, G. Kopidakis and S. Aubry,\\
 Phys.Rev.Lett. {\bf 85}, 550 (2000).
\item[7] O. Radulescu, T. Janssen , Phys. Rev. B{\bf 60}, 12737 (1999).
\item[8] L.A. Brussaard, A. Fasolino, T. Janssen, Phys. Rev. B {\bf 63}, 214302 (2001).
\item[9] T. Janssen, O. Radulescu and A.N. Rubtsov, subm. to Eur.Phys.J. B (2002).
\item[10] A.N. Rubtsov and T. Janssen, Phys. Rev. B {\bf 63}, 172101 (2001).
\item[11] R. Currat, E. Kats, I. Luk'yanchuk,
Eur. Phys. J. B {\bf 26}, 339 (2002)
\end{itemize}

\end{document}